\documentclass[preprint]{aastex631}

\usepackage{amsfonts}
\usepackage{amssymb}
\usepackage{amsthm}
\usepackage{amsmath}
\usepackage{alltt}
\usepackage{epstopdf}
\usepackage{graphicx}
\usepackage{float}
\usepackage{verbatim} 

\newcommand{\vel}{\boldsymbol{{u}}}

\submitjournal{ApJ}
\shortauthors{Korre \& Featherstone}

\begin{document}

\title{On the penetration of large-scale  flows into stellar radiative zones}

\author[0000-0002-0963-4881]{Lydia Korre}
\affiliation{Department of Applied Mathematics, University of Colorado, Boulder, CO 80309-0526, USA}
\author[0000-0002-2256-5884]{Nicholas A. Featherstone}
\affiliation{Southwest Research Institute, Department of the Space Studies, Boulder, CO 80302, USA}
\correspondingauthor{Lydia Korre}
\email{lydia.korre@colorado.edu}

\begin{abstract}
\noindent The propagation of meridional circulation below the base of the  convection zone  of low-mass stars may play a crucial role in the transport of angular momentum and  also significantly contribute to the transport of chemical species and magnetic fields within their stable radiative zone. We systematically study these large-scale mean flows by performing three-dimensional (3D) global numerical simulations in a spherical shell that consists of a convection zone {\color{black} (CZ)} overlying a stably stratified region. We find that  the meridional flows can penetrate  distances {\color{black} as large as $\sim 0.21r_o$ (where $r_o$ is the outer radius)} below the base of the convection zone, provided that the Eddington-Sweet timescale $t_{ES}$ is much shorter than the viscous timescale  $t_{\nu}$ as measured by the parameter $\sigma=(t_{ES}/t_{\nu})^{1/2}$. In the solar-like regime where $\sigma\lesssim 1$ in the upper radiative zone {\color{black} (RZ)}, we find that {\color{black} the angular momentum transport in the deep RZ} is determined primarily by the action of the Coriolis force on meridional flows. In contrast, in models run in the $\sigma> 1$ regime, the meridional flows become weaker and the  viscous effects dominate. We find that the penetration lengthscale $\delta_{MC}$ of these mean flows when $\sigma\lesssim 1$ is  proportional to $\sigma^{-0.22}$.  Our findings may provide a better understanding of the role of the meridional flows in the dynamics of the solar interior and inform future numerical studies  that are focused on capturing solar-like dynamics self-consistently.

\end{abstract}


\section{Introduction} 
\label{sec:intro}

  Helioseismic observations have provided us with the Sun’s rotational profile and have revealed that its radiative interior rotates almost uniformly while the  outer convective region rotates differentially, with the equator rotating faster than the poles \citep[e.g.,][]{Brown1989,Thompson2003,Howe2009}.  Differential rotation such as that observed in the Sun (i.e., with a rapidly-rotating equator) is thought to be maintained through anisotropic angular momentum transport arising in response to the tendency of convective cells to organize into columnar-like structures \citep[e.g.,][]{Zhang1992,Busse2002,Aurnou2007,Camisassa2022}.  
  
  The action of the Coriolis force on the differential rotation is such that near the rapidly-rotating equator, fluid motions are driven away from the rotation axis whereas  near the slowly rotating poles, the Coriolis force acts instead to push the flow toward the rotation axis.  The result of this process, known as ``gyroscopic pumping'', is to drive axisymmetric motions (the nominal meridional circulation) in the north/south and radially inward/outward directions.  Gyroscopic pumping  had been initially studied in the context of the Earth's  circulation and later discussed in the astrophysical context  for its role in the solar interior dynamics \citep{McIntyre2007,GA2009,GB2010,Wood2011,MH2011}.

In the near-surface layers, a variety of helioseismic and surface-feature-tracking techniques have revealed the Sun's meridional flow to be predominantly poleward within each hemisphere, with a characteristic speed of 10-20 m/s \citep[e.g.,][]{Komm1993,Svanada2007,Ulrich2010,Hathaway2022}.
Unlike the well-known solar rotational profile, however, there is no clear consensus on the structure of meridional circulation in the deep convection zone. In particular, substantial disagreement regarding the depth of the return flow, as well as the multi- or mono-cellular nature of the meridional flow, persists between measurements made using different instruments and helioseismic techniques  \citep[e.g.,][]{Zhao2013,Jackiewicz2015,Gizon2020}.   This disagreement increases with depth and, as a result, the question of the penetration of these mean flows below the base of the convection zone into the  underlying
radiative zone remains unanswered.  Meridional flows are nevertheless thought to play an important role in the dynamics that arise near the interface between the convection and radiative zones, mediating the mixing of chemical species \citep[e.g.,][]{Pin07}, the transport of angular momentum and the transport of magnetic fields \citep{GM98,McIntyre2007,Wood2011,
AG2013,WB18}.

Direct numerical simulations offer an alternative means to explore meridional flows below the base of the convection zone.  As with all such studies, however, care must be taken to ensure that the simulated system is operating in a parameter regime relevant to the physical system under consideration.  The sense of differential rotation realized in a spherical convection zone, for example, is known to depend on the ratio of the buoyancy and Coriolis forces.  When buoyancy is dominant, a so-called antisolar differential rotation with rapidly-rotating poles and a slowly-rotating equator develops \citep[e.g.,][]{Gastine14}.  For this reason, models designed to study the Sun are constructed such that the Coriolis force in dominant over buoyancy.

In a similar vein, the structure and penetration depth of meridional flows are thought to be determined by the competition of two timescales, namely the timescale associated with viscous diffusion and the timescale related to the advection of the gyroscopically-pumped meridional flows.  When viscous effects dominate, the meridional flows are exponentially damped in depth and confined to a small region below the inner convective boundary.  When viscous effects are negligible, gyroscopic pumping causes meridional flows to penetrate long distances below the base of the convection zone \citep{GB2008,GA2009,GB2010,WB12,AG2013}.

In solar-type stars, viscous effects are negligible near the convection zone-radiative zone interface, and in the case of the Sun, the timescale associated with the gyroscopically-pumped meridional flows is much faster  than that related to viscous diffusion in the upper part of the radiative region \citep[see Section \ref{sec:model}, and e.g.][]{WB12}.
A number of studies have examined angular momentum transport and the driving of meridional flow in this low-diffusivity regime.  Notably, the work of \citet{GA2009} and \citet{WB12} \citep[see, also][]{GB2008,GB2010}  demonstrated that when the proper ordering of timescales was respected, meridional flows penetrated large distances below the base of the convection zone.  
These calculations were carried out, however, either in axisymmetric spherical shells or in 3D Cartesian boxes.  That restricted geometry prevented the self-consistent development of differential rotation and meridional circulation, which require a fully 3D, spherical domain to self-consistently capture the mean flows and the convective motions that drive them.  On the other hand, those models that do account for these nonlinear and geometric effects were run in a regime where viscous effects were dominant.  For instance, \cite{Brun2006} and \cite{Strugarek2011}   found in fully nonlinear simulations employing 3D, spherical geometry that the angular momentum transport was dominated by viscosity within the bulk of the RZ and the meridional circulation did not propagate substantially beyond the base of the convection zone. Recently, \cite{Matilsky22} also ran 3D magnetohydrodynamic (MHD) global simulations to study the self-consistent formation of the tachocline due to dynamo action within the stable radiative zone. Their simulations were run in the non-solar regime and consequently they also found that viscous diffusion played a dominant role in their dynamics.

 In this paper, we seek to bridge the gap between these two approaches.  Our goal is to understand the dynamics occurring within the convection zone-radiative zone (CZ-RZ) interface, in a fully 3D, spherical geometry, by exploring a range of parameter regimes in which either viscous or gyroscopic pumping effects dominate.  In Section \ref{sec:model}, we describe our two-zone spherical shell formulation and provide the set of anelastic Navier-Stokes equations along with the initial conditions, input parameters, and the boundary conditions used in our simulations. In Section \ref{sec:results}, we present our results related to the differential rotation and meridional circulation profiles within and below the CZ and their dependence on the input parameters. We also discuss angular momentum transport within the different parameter regimes and compare the time evolution of the corresponding fluxes among the different cases. Finally in Section \ref{sec:disc}, we provide a summary of our findings along with a  discussion of their relation to other studies and  the implications of these results for the solar interior dynamics. 

\section{Model Formulation} 
\label{sec:model}
\subsection{Dimensional Equations}
We are interested in exploring the dynamics related to the global, large-scale flows arising in rotating overshooting convection by accounting for a two-layered system with a stably stratified region underlying a convective region. We assume a fixed aspect ratio  $r_i/r_o=0.45$ where $r_i$ is the inner radius  and $r_o$ is the outer radius of the spherical shell while the  depth of the convective region is given by $L=r_o-r_{c}=0.2408r_o$, where $r_{c}=0.7592r_o$ is the radius at the inner boundary of the CZ. The depth $L$ is chosen such that it corresponds to the  solar convection zone from $\sim 0.7187R_{\odot}$ to {\color{black}$\sim  0.9041R_{\odot}-0.9695R_{\odot}$}, since we aim to focus on the deep interior dynamics and not the radiative transfer processes occurring near and at the  outer solar convective boundary. Consequently,  we employ the anelastic approximation which assumes small perturbations of the thermodynamic variables compared with their mean  while it also filters out the sound waves \citep{anelasticG,anelasticGG}. We solve the 3D Navier-Stokes equations  under the anelastic and Lantz-Braginsky-Roberts approximations \citep{anelastic1,anelastic2} with the latter being  exact where the reference state is adiabatic.

Then, the dimensional Navier-Stokes equations become
\begin{equation}
\label{eq:momeq}
\dfrac{\partial{\vel}}{\partial{t}} +\vel\cdot\nabla\vel+2\Omega_o \hat{z}\times\vel=\dfrac{g(r)}{c_p}S\mathbf{\hat{r}}-\nabla(P/\bar{\rho})+\dfrac{1}{\bar{\rho}}\nabla\cdot \mathbf{D},
\end{equation}
\begin{equation}
\label{eq:conteq}
\nabla\cdot(\bar{\rho}\vel)=0,
\end{equation}

\begin{equation}
\label{eq:energyeq}
\bar{\rho}\bar{T}\left(\dfrac{\partial{S}}{\partial{t}}+\vel\cdot\nabla S +u_r\dfrac{d\bar{S}}{dr}\right)=\nabla\cdot(\bar{\rho}\bar{T}\kappa\nabla S)+Q+\mathbf{\Phi},
\end{equation}
where $\vel=(u_r,u_{\theta},u_{\phi})$ is the velocity field,  $P$ is the pressure, $\bar{\rho}$ is the reference density,  $\bar{T}$ is  the reference  temperature,  $d\bar{S}/dr$ is the background entropy gradient, $S$ corresponds to the entropy perturbations about the reference state, $\Omega_o$ is the {\color{black} stellar mean frame  rotation rate (for the Sun, this corresponds to $\Omega_0=2.87\cdot 10^{-6}$ rad/s)}, $g(r)$ is the gravity (where $g\propto 1/r^2$), and $c_p$ is the specific heat at constant pressure. We assume that the  viscosity $\nu$ and the thermal diffusivity $\kappa$  are  constant for simplicity. The viscous stress tensor $\mathbf{D}$ is given by
\begin{equation}
\mathbf{D}=2\bar{\rho}\nu(e_{ij}-\dfrac{1}{3}\nabla\cdot\vel),
\end{equation}
and the viscous heating is denoted by 
\begin{equation}
\mathbf{\Phi}=2\bar{\rho}\nu[e_{ij}e_{ij}-\dfrac{1}{3}(\nabla\cdot\vel)^2],
\end{equation}   
where $e_{ij}$ is the strain rate tensor. We adopt an internal heating term $Q$  that is  similar  to the one formed in Model S \citep[see e.g.,][]{ModelS} by making it proportional to the reference pressure $\bar{P}$ \citep[for more details see][]{KF21}. The internal heating satisfies $Q(r)=-\nabla\cdot F_{rad}$, where $F_{rad}$ is the radiative flux in the system  given by
\begin{equation}
\label{eq:Frad}
F_{rad}(r)=\dfrac{1}{r^2}\int_{r}^{r_o}Q(r')r'^2dr'.
\end{equation}
The equation of state is
\begin{equation}
\label{eqstate}
\dfrac{\rho}{\bar{\rho}}=\dfrac{P}{\bar{P}}-\dfrac{T}{\bar{T}}=\dfrac{P}{\gamma\bar{P}}-\dfrac{S}{c_p},
\end{equation}
assuming the ideal gas law
\begin{equation}
\label{eq:gaslaw}
\bar{P}=\Re\bar{\rho}\bar{T},
\end{equation}
where $\rho$  and  $T$ characterize  respectively the density and temperature perturbations about the reference state, $\Re$ is the gas constant, and $\gamma=c_p/c_v$, where $c_v$ is the specific heat at constant volume.

\par To create our two-zone spherical shell, we  choose a profile of $d\bar{S}/dr$  that satisfies an adiabatic polytropic solution with $\gamma=5/3$  in the convective region $r_{c}\leq r\leq r_o$ and that forms  a stably stratified region for $r_i\leq r< r_{c}$.  

We also designate the number of density scale-heights in the convection zone through the parameter $N_{\rho}=\ln(\bar{\rho}(r_c))/\bar{\rho}(r_o))$ (see Appendix of \cite{KF21} for details). 
Then, we integrate $d\bar{S}/dr$ from the bottom of the convection zone $r_c$ inward which results in a solution for  $\bar{S}$ with an integration constant that matches the solution to the value of $\bar{S}$ at $r_c$.  \\
Given that the entropy for a monatomic ideal gas is   $\bar{S}=\ln(\bar{P}^{1/\gamma}/\bar{\rho})$, we can differentiate $\bar{S}$ with respect to $r$, and  apply the hydrostatic balance equation ${\partial \bar{P}}/{\partial r}=-\bar{\rho} {g}$. Then, we obtain
\begin{equation}
\label{dsdrprofile}
\dfrac{\bar{\rho}}{c_p}\dfrac{{d\bar{S}}}{dr}+\dfrac{{g}}{\gamma}\exp({{-\gamma \bar{S}}/{c_p}})\bar{\rho}^{(2-\gamma)}+\dfrac{d\bar{\rho}}{dr}=0.
\end{equation}
This allows us to numerically solve for the reference density profile $\bar{\rho}(r)$ of any chosen background entropy gradient profile. 

\subsection{Nondimensional Equations}
\label{sec:ND}
\begin{figure}[ht]
     \centering
     \includegraphics[scale=.45]{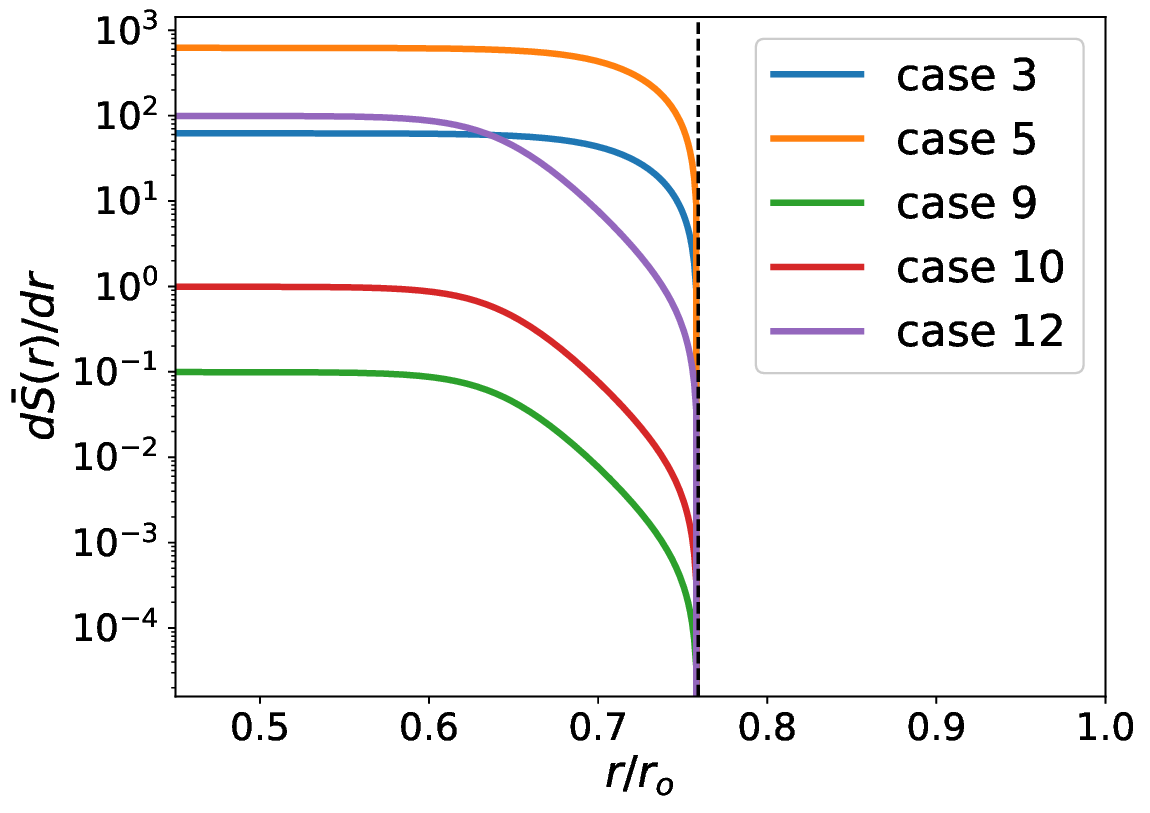}
     \caption{Profiles of the nondimensional background entropy gradient $d\bar{S}(r)/dr$  versus the radius $r/r_o$ for five representative runs with different $A_S$, and/or $r_b$ and/or $d$ (see Eq. (\ref{eq:dsdr})). The black dashed vertical  like marks the transition from the adiabatic convection zone (where $d\bar{S}(r)/dr=0$) to the subadiabatic layer (where $d\bar{S}(r)/dr>0$).}
     \label{fig:ds}
 \end{figure}

We nondimensionalize equations (\ref{eq:momeq})-(\ref{eq:energyeq}) using the depth of the convection zone $[l]=L$ as the lengthscale, the velocity scale $[u]=\nu/L$ and the viscous timescale $[t]=L^2/\nu$. Throughout the paper, we define the volume average of a quantity $q(r,\theta,\phi)$ in the CZ as 
\begin{equation}
\label{eq:volCZ}
{{q}_{cz}=\dfrac{\int_{r_{c}}^{r_o}\int_0^{2\pi}\int_0^{\pi} q(r,\theta,\phi) r^2\sin\theta d\theta d\phi dr}{\int_{r_{c}}^{r_o}\int_0^{2\pi}\int_0^{\pi}  r^2\sin\theta d\theta d\phi dr}},
\end{equation}
 the time and spherical average of a quantity $q$ as 
\begin{equation}
\label{eq:qav}
\tilde{q}(r)=\dfrac{1}{4\pi(t_2-t_1)}\int_{t_1}^{t_2}\int_0^{2\pi}\int_0^{\pi}q(r,\theta,\phi,t)\sin\theta d\theta d\phi dt,
\end{equation}
{\color{black}
and the time and azimuthal average of a quantity $q$ as
\begin{equation}
\label{eq:qaz}
\langle q(r,\theta) \rangle=\dfrac{1}{2\pi(t_2-t_1)}\int_{t_1}^{t_2}\int_0^{2\pi}q(r,\theta,\phi,t)d\phi dt.
\end{equation}
}
For the density and temperature scales, we use
$[\rho]=\bar{\rho}_{cz}$, and $[T]=\bar{T}_{cz}$, respectively.
For  the entropy perturbations $S$, we choose a scaling associated with the thermal energy flux $F$ such that $[S]={L{F}_{cz}}/({{\bar{\rho}}_{cz}{\bar{T}}_{cz}\kappa})$, where $F(r)=\int_{r_i}^{r}Q(r')r'^2dr'/{r^2}$.
We  determine our selected nondimensional $d\bar{S}/dr$ profile in the RZ using the function
\begin{equation}
\label{eq:dsdr}
\dfrac{d\bar{S}}{dr}=\dfrac{A_S}{2}\left(1-\tanh\left(\dfrac{r-r_{b}}{d}\right)\right),
\end{equation}
where we vary $A_S$, $r_{b}$ and $d$. In Figure \ref{fig:ds}, we show the profiles of $d\bar{S}(r)/dr$ versus $r/r_o$ for five typical runs. We note that for larger values of $A_S$, we obtain a more stably stratified RZ while the different values of $r_b$ and $d$ control the smoothness of the transition width between the two layers.

The nondimensional reference  density $\bar{\rho}$, reference temperature $\bar{T}$ and gravity are equivalent to the dimensional reference density, reference temperature and gravity divided by  $\bar{\rho}_{cz}$, $\bar{T}_{cz}$, and $g_{cz}$, respectively. The  nondimensional anelastic Navier-Stokes equations become
\begin{equation}
\label{eq:NDmom}
\dfrac{\partial{\vel}}{\partial{t}} +\vel\cdot\nabla\vel+\dfrac{2}{\rm{Ek}}\hat{z}\times\vel=\dfrac{\rm{Ra}}{\rm{Pr}}g(r)S\mathbf{\hat{r}}-\nabla(P/\bar{\rho})+\dfrac{1}{\bar{\rho}}\nabla\cdot \mathbf{D},
\end{equation}

\begin{equation}
\label{eq:NDconteq}
\nabla\cdot(\bar{\rho}\vel)=0,
\end{equation}

\begin{equation}
\label{eq:NDenergyeq}
\bar{\rho}\bar{T}\left(\dfrac{\partial{S}}{\partial{t}}+\vel\cdot\nabla S +u_r\dfrac{d\bar{S}}{dr}\right)=\dfrac{1}{\rm{Pr}}\nabla\cdot(\bar{\rho}\bar{T}\nabla S)+\dfrac{1}{\rm{Pr}}Q_{nd}+\dfrac{\rm{Di Pr}}{\rm{Ra}}\mathbf{\Phi},
\end{equation}
where $Q_{nd}=LQ/{{F}_{cz}}$ is  the nondimensional internal heating function  \citep[for details, see][]{KF21}. We note that $\vel=\vel_m+\vel_f$, where $\vel_m$ is the mean, axisymmetric part of the velocity corresponding to the large-scale
motions related to rotation and $\vel_f$ is the fluctuating
component of the velocity field associated with the convective motions. 

In all that follows, all the variables and parameters are now
 nondimensional and this nondimensionalization introduces the flux Rayleigh number Ra, the Prandtl number Pr,  the Ekman number Ek and the  dissipation number Di, which are defined respectively as
\begin{equation}
\label{eq:RaPrEDi}
{\rm{Ra}}=\dfrac{{g}_{cz}{F}_{cz}L^4}{c_p\bar{\rho}_{cz}\bar{T}_{cz}\kappa^2\nu},\quad
{\rm{Pr}}=\dfrac{\nu}{\kappa},\quad
{\rm{Ek}}=\dfrac{\nu}{\Omega_oL^2},\quad\text{and}\quad 
{\rm{Di}}=\dfrac{{g}_{cz}L}{c_p\bar{T}_{cz}}.
\end{equation}
We note that although  Ra, Pr and Ek are free input parameters,  Di is not as it depends on the chosen $N_{\rho}$ and  accounts for the fact that we have both a thermal scale and a kinetic scale. Thus, Di appears in the thermal equation to account for the viscous heating term, which is intrinsically kinetic.

We have run a series of 3D numerical simulations solving equations (\ref{eq:NDmom})-(\ref{eq:NDenergyeq}) using the open-source convection code Rayleigh   \citep{Featherstone16a, Matsui16,Rayleighcode}. We vary the background entropy gradient $d\bar{S}/dr$ as described above, the Rayleigh number, the Ekman number, the number of density scale-heights in the convection zone $N_{\rho}$ while we keep the Prandtl number fixed at Pr $=1$ in order to avoid the excitation of unphysical modes that are known to appear in rapidly rotating anelastic convection systems for  values of Pr $< 1$ \citep[e.g.,][]{Calkins15}.

We provide the parameters used in our simulations in Table \ref{tab:table}. We also report on the convective Rossby number Ro$_c=\sqrt{\rm Ra Ek^2/(4 Pr)}$ and the output Rossby number  defined as Ro $=U/(2L\Omega_0)=$ ReEk/2,
which is the ratio of inertial forces to the Coriolis force. Here, the first expression involves dimensional quantities where $U = u_{rms}\nu/L$ is
a typical dimensional velocity in the CZ and where for the value of $u_{rms}$, we
use the total (fluctuating part and mean part) velocity
averaged over the CZ and weighted by density. The second
expression includes nondimensional quantities where
Re$ =UL/\nu$  is the Reynolds number extracted from the
simulations while the 
P\'{e}clet number Pe is Pe = PrRe = Re in our Pr = 1 runs (see Table \ref{tab:table}).

We use impenetrable and stress-free boundary conditions for the velocity while for the entropy perturbations, we assume $\partial S/\partial r|_{r_i}=0$ at the inner boundary and  $S|_{r_o}=0$ at the outer boundary.  
Each simulation is evolved  from a zero initial
velocity and small-amplitude perturbations in the entropy field
until a statistically stationary and thermally equilibrated state is achieved.

\subsection{Choice of Input Parameters}

\begin{figure}[ht]
     \centering
     \includegraphics[scale=0.3]{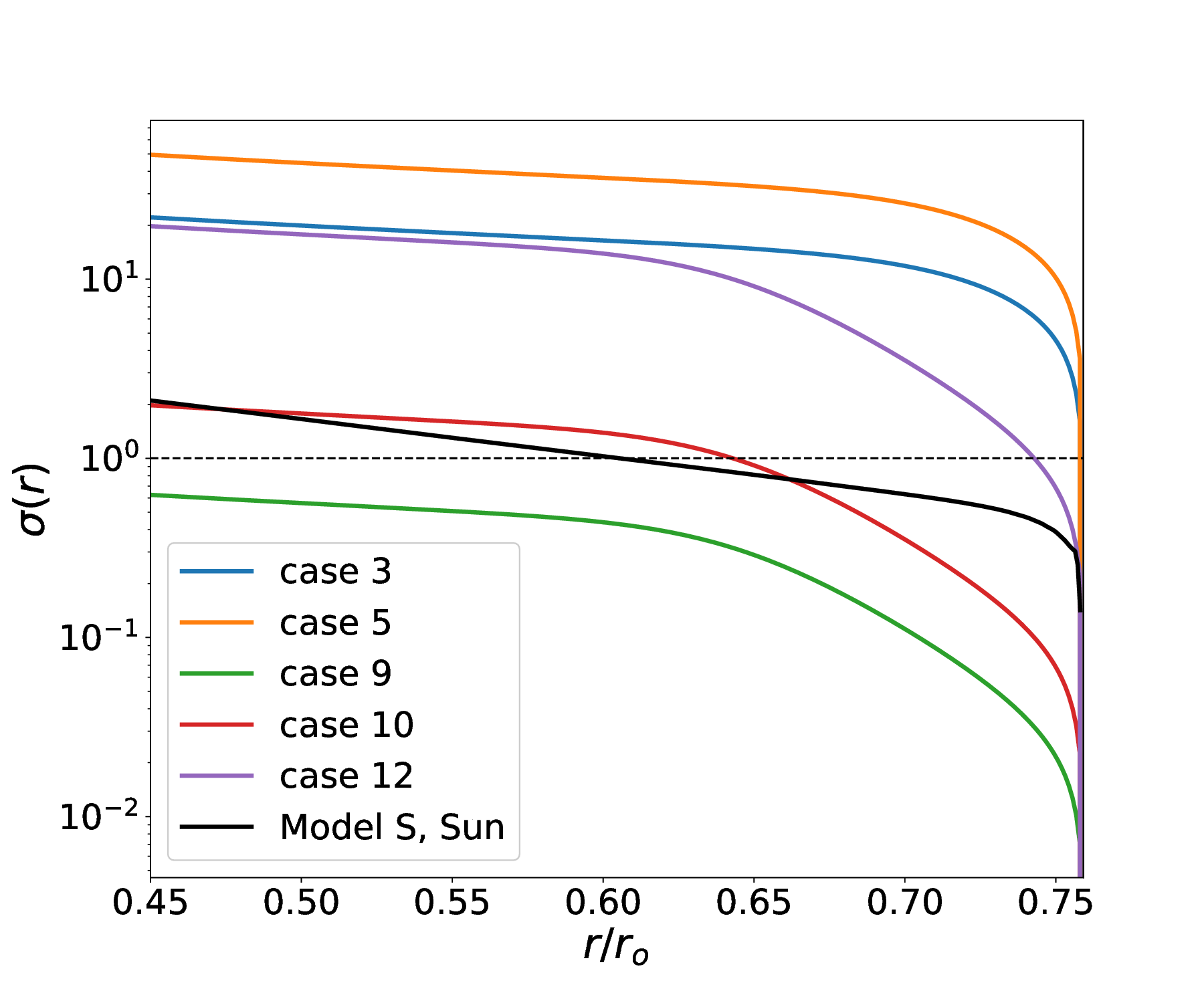}
     \caption{Profiles of $\sigma(r)$ versus $r/r_o$ for the five runs with the $d\bar{S}/dr$ profiles presented in Figure \ref{fig:ds}. The horizontal dashed black line corresponds to $\sigma(r)=1$, while the black line corresponds to the computed solar $\sigma(r)$ profile with the base of the solar convection zone moved by $0.046r_o$ to coincide with  the base of the CZ of the spherical shell (see text for details). }
     \label{fig:sigma1}
 \end{figure}

As discussed in Section \ref{sec:intro}, in order for the meridional circulation to penetrate below the base of the CZ more deeply into the stable region, the correct ordering of timescales needs to be achieved in numerical simulations even though the latter cannot account for real stellar values. The gyroscopically-pumped meridional circulation operates on a timescale of the order of the  Eddington-Sweet timescale $t_{ES}=N^2L^2/(\Omega_0^2\kappa)$ \citep[see, e.g.][]{SZ92,WB12} which is shorter than the viscous timescale $t_{\nu}=L^2/\nu$ in the upper solar radiative zone (Fig. \ref{fig:sigma1}). The ordering of timescales, namely $t_{ES}<t_{\nu}$,    can be expressed via the parameter 
$\sigma=({t_{ES}}/{t_{\nu}})^{1/2}=\sqrt{\nu/\kappa}N/\Omega_0$ \citep{GA2009,GB2010,WB12}.
In our nondimensional formulation, $\sigma$ is written as
\begin{equation}
\label{eq:sigmaeq}
\sigma(r)=\sqrt{\rm Pr}N(r){\rm Ek},
\end{equation}
where $N(r)=\sqrt{({\rm Ra}/{\rm Pr})g ({d\bar{S}(r)}/{dr})}$ is the Brunt-V\"{a}is\"{a}l\"{a} frequency.

 We are interested in systematically investigating the dependence of the meridional flows on   $\sigma$ within a 3D spherical shell that self-consistently accounts for large-scale mean flows. To achieve this,  we vary $\sigma(r)$ below the base of the convection zone. In our Pr $=1$ simulations, we can do so by varying Ek and/or $N(r)$ through Ra and/or $d\bar{S}(r)/dr$. We note, however, that in our runs the variation in $\sigma(r)$ is mostly a result of the  different $d\bar{S}/dr$ profiles chosen.  
 
 In Table \ref{tab:table}, we show all the  cases considered in this study by reporting $\sigma_{ov}$ which is $\sigma$ evaluated at the radius down to which the convective motions overshoot in each case. That is at $r_c-\delta_G r_o$, where $\delta_G=\delta_{Gf}/r_o$ is the overshoot lengthscale (for more details on how we calculate this, see Section 3 from \cite{KF21}). In \citet{KF21}, we found that a convective Rossby number Ro$_c\approx 0.5$ leads to solar-like rotation with the equator rotating faster than the poles in the convective region. Since we are interested in {\color{black} low-mass} stars, we vary our input parameters such that we are in the Ro$_c\approx 0.5$ regime and study the influence of $\sigma$ on the mean flow dynamics.

In Figure \ref{fig:sigma1}, we show the function $\sigma(r)$ versus the radius $r/r_o$ below the base of the convection zone for the same five representative runs illustrated in Figure \ref{fig:ds}.  We also show the solar $\sigma(r)$ profile calculated from Model S \citep{ModelS}, where we have moved the solar convection zone by $0.046r_o$ to match the base of the CZ of our spherical shell so that all the  $\sigma(r)$ profiles can be directly comparable in the same plot. For the calculation of the solar $\sigma(r)$, we have used the density, temperature, pressure, Brunt-V\"{a}is\"{a}l\"{a} frequency, specific heat at constant pressure and gravity of Model S  while we have calculated the thermal diffusivity and viscosity (and as a result the Prandtl number) using the formulae given in \citet{GG08} \citep[see also][]{Gough07}. We see that  case 9 has a smaller than one $\sigma(r)$ profile within the whole RZ, while  in case 10, $\sigma(r)$ is overall small but smaller than unity only near  the upper part of the radiative zone, similarly to the actual solar case.  Case 12 possesses a  $\sigma<1$ profile near the upper stable layer unlike cases 3 and 5 which both have  $\sigma>>1$ across  the whole RZ. By illustrating different diagnostics of  these five typical cases that span a wide range of $\sigma_{ov}$ values, in the following sections, we will explore the dynamics associated with the large-scale mean flows, and their dependence on $\sigma$.

\begin{deluxetable*}{ccccccccccccccc}
\label{tab:table}
\centering
\tablecolumns{10}
\tablewidth{0pc}
{
\tablecaption{Input and output parameters of the simulations}
\tablehead{
\colhead{Case} & \colhead{$N_{\rho}$} & \colhead{Ra} & 
& \colhead{Ek}   &  \colhead{Ro$_c$} & $N_r\times N_{\theta}\times N_{\phi}$ & \colhead{Ro} & \colhead{$\delta_G$} & \colhead{Re} & \colhead{$\delta_{MC}$} & \colhead{$\sigma_{ov}$}}
\startdata
1 & 2 & $10^6$ & 
&   0.001  &  0.5 & 192$\times$1104$\times$2208 &   0.054 & 0.023 &   108.5 &  0.096 & 1.9\\
2 & 2 & $10^5$ & 
&   0.0032 &  0.51 & 192$\times$1104$\times$2208 &   0.052 & 0.011 &   32.3 & 0.054  & 3.5\\
3 & 2 & $10^5$ & 
&   0.0032 &  0.51 & 192$\times$1104$\times$2208 &   0.051 & 0.010 &   32.1 & 0.037 & 4.7  \\
4 & 2 & $10^5$ & 
&   0.0032 &  0.51 & 192$\times$1104$\times$2208 &   0.051 & 0.0094 &   31.9 & 0.0275 & 7.4  \\
5 & 2 & $10^5$ & 
&   0.0032 &  0.51 & 192$\times$1104$\times$2208 &   0.051 & 0.0089 &   31.8 & 0.022 & 9.8 \\
6 & 3 & $10^6$ & 
&   0.001   & 0.5 & 192$\times$1104$\times$2208 & 0.0432 & 0.021 & 86.4 & 0.123 & 0.38 \\
7 & 3 & $5\cdot 10^5$ & 
&   0.0015  &  0.53 & 192$\times$1104$\times$2208 & 0.046   & 0.019  & 61.9 & 0.176 & 0.09\\
8 & 3 & $10^5$ & 
&   0.0032  &  0.51 & 192$\times$1104$\times$2208 &   0.0435 & 0.011 &   27.2 & 0.056 & 3.5 \\
9 & 4 & $10^5$ & 
&   0.0032  &  0.51 & 192$\times$1104$\times$2208 & 0.0389   & 0.017  & 24.3 & 0.213 & 0.032 \\
10 & 4 & $10^5$ & 
&   0.0032   &  0.51 & 192$\times$1104$\times$2208 &  0.0387 &  0.017  & 24.2 & 0.164 & 0.1 \\
11 & 4 & $10^5$ & 
&   0.0032 &  0.51 & 192$\times$1104$\times$2208 & 0.0382  &  0.017 &  23.9 &  0.131  & 0.32 \\
12 & 4 & $10^5$ & 
&   0.0032  &  0.51 & 192$\times$1104$\times$2208 &  0.0377 & 0.015   &  23.5 & 0.1 & 1.1 \\
\enddata
\tablecomments{Columns $2-5$ indicate the input  parameters, column $6$ provides the resolution  and columns $7-11$ report on the output parameters of the simulations.}}
\end{deluxetable*}

\section{Results}
\label{sec:results}
\subsection{Differential Rotation Profiles}
\begin{figure*}[t!]
     \centering
     \includegraphics[scale=.55]{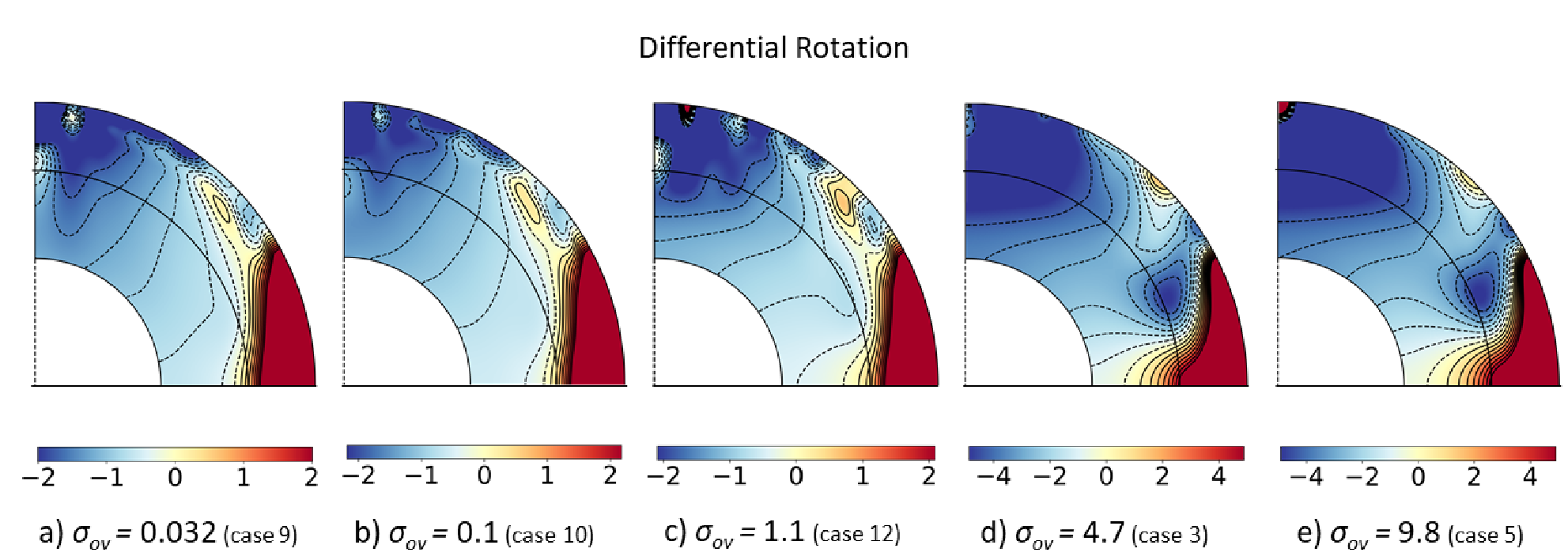}
     \caption{Differential rotation profiles {\color{black} $\langle u_{\phi}\rangle/(r\sin\theta)$} for the five representative runs spanning a wide range of $\sigma_{ov}$ from  a) $\sigma_{ov}=0.032$ up to e) $\sigma_{ov}=9.8$. For decreasing values of $\sigma_{ov}$, the differential rotation is more confined to the upper radiative zone within the equatorial region.}
     \label{fig:diffrot}
 \end{figure*}

We begin our investigation by focusing on the rotational profiles of our simulations and the mean flows associated with the differential rotation within and below the CZ.
In Figure \ref{fig:diffrot}, we illustrate profiles of the differential rotation $\langle u_{\phi}\rangle/(r\sin\theta)$, for the five representative cases shown in Figure \ref{fig:sigma1}. We verify that for Ro$_c\approx 0.5$, the simulations have a solar-like rotation whereby the equatorial region rotates more rapidly than the polar regions throughout the convection zone.

We do note that, while these models incorporate a region of overshoot, contours of isorotation are largely parallel to the rotation axis, as opposed to tilted in the radial direction as observed in the Sun \citep[e.g.,][]{Howe2005}.  This ``tilting'' of the isorotation contours has been posited to arise in response to strong latitudinal temperature variations established in the tachocline region \citep{Rempel2005, Miesch2006}.  As in \citet{KF21}, we observe only very weak deviations from cylindrical isorotation contours, in spite of the fact that a region of overshoot has been included in these simulations.  This is possibly because the thermal-wind balance required to induce deviations from cylindrical isorotation is instead produced in response to uniform thermal flux at the upper boundary, which is not imposed for the models considered here \citep[e.g.,][]{Matilsky20}.

 \par None of the models illustrated in Figure \ref{fig:diffrot} possess a true tachocline of shear at the base of the convection zone.  For all values of $\sigma_{ov}$, differential rotation profiles established in the convection zone are reflected to some extent in the radiative zone.  This behavior is largely in accord with the widely-accepted consensus that additional physics, such as  magnetic fields (either primordial, or dynamo, or both) are needed to reproduce the Sun's rotation and the self-consistent emergence of a tachocline \citep{GM98, AG2013, WB18, Matilsky22}. 
 
\begin{figure*}[ht]
     \centering
  \includegraphics[scale=.5]{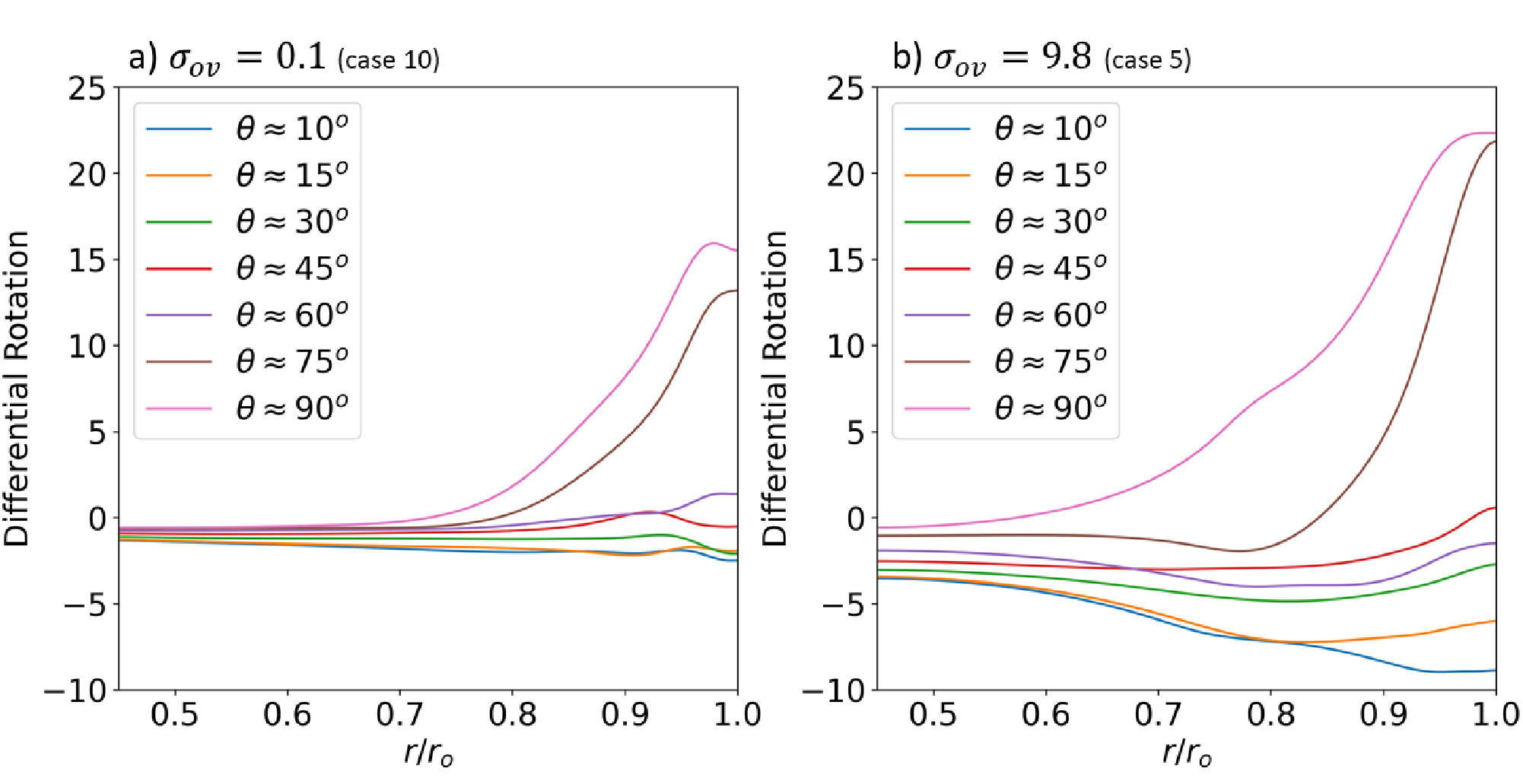}
    \caption{Profiles of the time- and azimuthally-averaged differential rotation profiles  versus the radius $r/r_o$ at different colatitudes $\theta$ for a) case 10 with $\sigma_{ov}=0.1$ and b) case 5 with $\sigma_{ov}=9.8$.}
    \label{fig:Omegatheta}
    \end{figure*}

 We note, however, that for the low-$\sigma_{ov}$ cases, differential rotation in the equatorial region is considerably more confined to the upper radiative zone than  in the high-$\sigma_{ov}$ regime runs. This can be seen more  clearly  in Figure \ref{fig:Omegatheta} where we illustrate the radial dependence of the differential rotation at different colatitudes for the case with   $\sigma_{ov}=0.1$ and the case with   $\sigma_{ov}=9.8$. We verify that  the profiles   are quite different between the two cases, with the low-$\sigma_{ov}$ run presenting a smaller variation of differential rotation below the base of the CZ and a profile at $\theta\approx 90^o$ that drops more rapidly beyond $r_c$, unlike the high-$\sigma_{ov}$ case. As we discuss in Section \ref{sec:AMT}, in the systems with $\sigma_{ov}\lesssim 1.1$, it is advection of angular momentum by meridional circulation that halts the spread of differential rotation that is otherwise observed in the equatorial regions of models with  $\sigma_{ov}>>1$, where viscous effects dominate.  To see why this might be the case, we now examine the accompanying meridional circulations established in these different models.

\subsection{Meridional Flows Below the Base of the Convection Zone}
\begin{figure*}[ht]
     \centering
     \includegraphics[scale=.55]{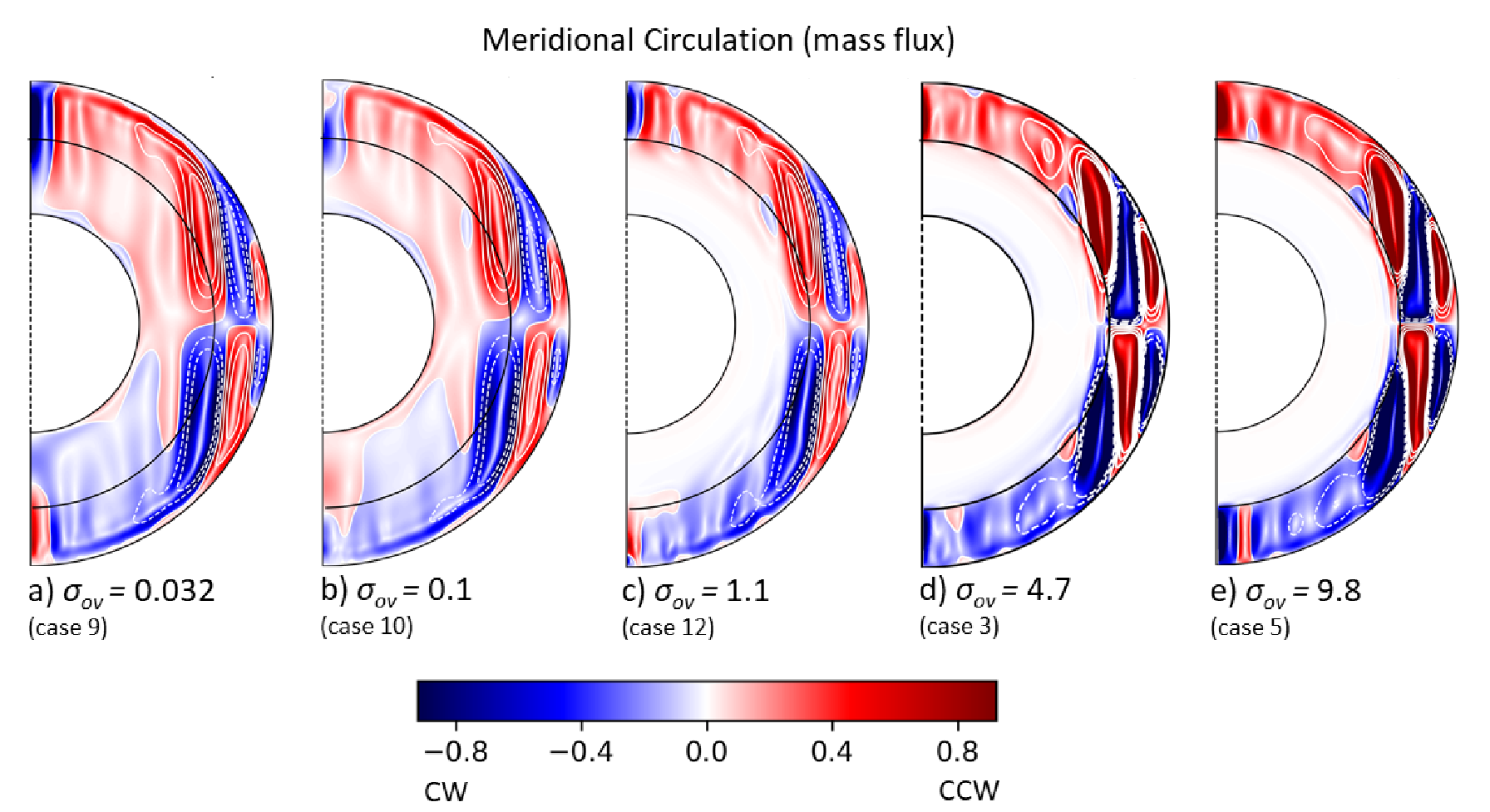}
     \caption{Time- and azimuthally-averaged profiles of the meridional circulation streamlines with underlying contours of the mass flux where blue color corresponds to clockwise motion (CW), and red color to counterclockwise motion (CCW) for the five typical runs spanning values of $\sigma_{ov}$ from a) 0.032 to e) 9.8. The meridional flows present multiple cells within the convective region and for the cases where $\sigma_{ov}<1$ (a) and b)), the flows penetrate more deeply into the stable layer below. In panel c) where $\sigma_{ov}=1.1$, there is still substantial propagation of the meridional circulation below the base of the CZ while in panels d) and e) where $\sigma_{ov}>1$, the mean flows only travel a small distance beyond the inner convective boundary. }
     \label{fig:MCflows}
 \end{figure*}

We present profiles of the meridional circulation  in Figure \ref{fig:MCflows}.  There, we plot the time- and
azimuthally-averaged meridional circulation streamlines overlying the mass flux function $\pm\sqrt{\bar{\rho}u_r^2+\bar{\rho}u_{\theta}^2}$ for those models illustrated in Figure \ref{fig:diffrot}.  The mass flux has been colored to indicate the handedness of the flow, such that clockwise circulations are blue, and counterclockwise circulations are red. 
Within the convection zone, the meridional circulation is similar across all models, with only minor differences among their amplitudes.  Notably, all systems exhibit multiple circulation cells within each hemisphere, as expected for the Ro$_c<1$ regime in which these models were calculated \citep[e.g.,][]{FM2015,KF21,Camisassa2022}. 

It is below the base of the convection zone where significant differences in the meridional flow morphology arise.  For the low-$\sigma_{ov}$ models 9 and 10, the mean flows penetrate deeply into the radiative zone, even reaching the lower boundary of the simulation domain, although their amplitude is much diminished there.  Models 3 and 5 possess values of $\sigma$ that are  larger than unity everywhere in the radiative zone,  and their resulting meridional flows are confined to a small depth just below the base of the CZ.  The intermediate case 12, which possesses  $\sigma(r)<1$ only in the upper radiative zone (and a $\sigma_{ov}$ of 1.1) has meridional flows that  travel well below the base of the CZ,  beyond the overshoot region but their amplitude rapidly decays in the deeper stable zone.

These results qualitatively support the earlier theoretical arguments and restricted-geometry models discussed in e.g. \cite{GB2008}, \cite{GA2009}, \cite{WB12}, and \cite{AG2013}.  When $\sigma>>1$, in our runs $N(r)$ is large, namely the thermal stratification in the RZ is stronger and   the meridional flows  are viscously-damped before they can extend to large depths below the base of the CZ unlike in the cases with $\sigma<1$ where the meridional flows are stronger and  are able to propagate longer distances into the RZ and even span the whole stable layer \citep{GB2008,GB2010}. 

\begin{figure*}[ht]
     \centering
     \includegraphics[scale=.55]{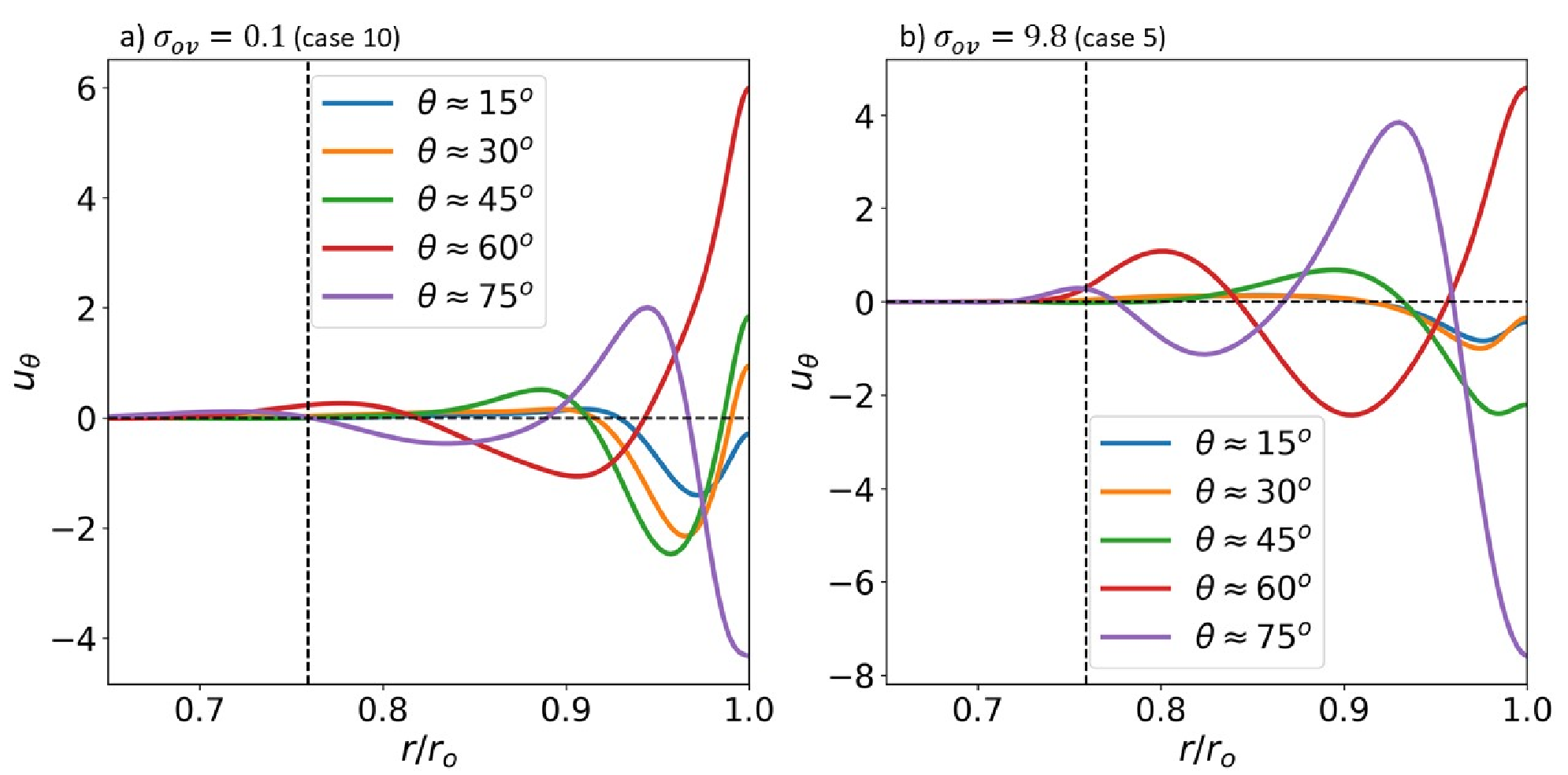}
     \caption{\color{black}{Meridional circulation profiles at different $\theta$. Time- and azimuthally-averaged profiles of $u_{\theta}$  plotted from $0.65r_o$ to $r_o$ at different colatitudes. The profiles illustrate the meridional flows in and below the base of the CZ for a) $\sigma_{ov}=0.1$ (case 10) and b) $\sigma_{ov}=9.8$ (case 5).} }
     \label{fig:MCtheta}
 \end{figure*}

{\color{black} A more quantitative picture of the meridional circulation can be seen in Figure \ref{fig:MCtheta}. There, the time- and azimuthally-averaged  velocity $u_{\theta}$ is shown as a function of radius at different colatitudes for the solar-like case 10 with $\sigma_{ov}=0.1$ and the non-solar case 5 with $\sigma_{ov}=9.8$. We observe that the meridional circulation is roughly similar in amplitude and structure at all $\theta$ in the CZ for both cases, with the multicellular profiles seen by the  change of sign of $u_{\theta}$ in the convective region. 
We note that the meridional circulation profiles are unlike those obtained from observations near the solar surface due to diffusive effects in the  boundary layer associated with the outer boundary condition employed in our numerical simulations 
\citep[for a more detailed discussion on this topic, see,][]{Fuentes_2024}.}

\begin{figure*}[ht]
     \centering
     \includegraphics[scale=.55]{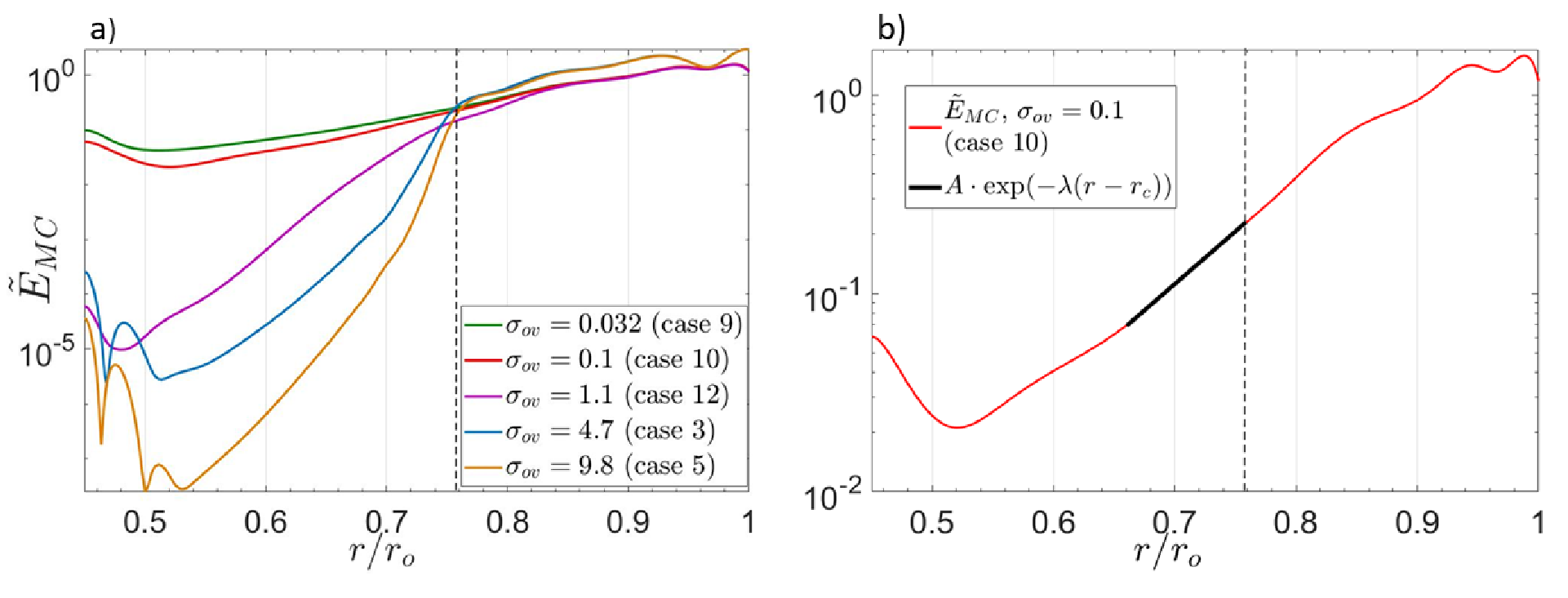}
     \caption{Time- and spherically-averaged profiles of the kinetic energy in the meridional flows versus $r/r_o$. a) Profiles for the five representative runs. $\tilde{E}_{MC}$ is almost the same for all runs within the CZ, but changes for the different $\sigma_{ov}$ cases in the RZ. For increasing values of $\sigma_{ov}$, $\tilde{E}_{MC}$ decreases much faster below the base of the CZ compared with the smaller $\sigma_{ov}$ cases. b) $\tilde{E}_{MC}$ profile for case 10 with $\sigma_{ov}=0.1$. An exponential function (see Eq. (\ref{eq:fitKE})) is fitted to the kinetic energy data below the base of the CZ down to the point where $\tilde{E}_{MC}$ is well fit. The computed decay rate $\lambda$ for each simulation provides a penetration lengthscale $\delta_{MC}=2/\lambda$ for all runs shown in Table \ref{tab:table}. The inner dashed black line corresponds to the base of the CZ in both panels.}
     \label{fig:KE_MC}
 \end{figure*}

These trends in the meridional flow behavior are also evident in the radial profiles of the time- and spherically-averaged kinetic energy related to the meridional circulation, $\tilde{E}_{MC}$, which we define as
\begin{equation}
\label{eq:KE_MC}
\tilde{E}_{MC}(r)=\frac{1}{2}\bar{\rho}(\widetilde{u_{m,r}^2}+\widetilde{u_{m,\theta}^2}).
\end{equation}
We present profiles of $\tilde{E}_{MC}$ for each  one of our representative cases in Figure \ref{fig:KE_MC}a.  We observe that $\tilde{E}_{MC}(r)$ is approximately the same in the convection zone for all five cases but quite different within the RZ.  Indeed, $\tilde{E}_{MC}(r)$ decays faster beyond the inner convective boundary as $\sigma$ becomes larger there for the cases with $\sigma_{ov}>1$. More specifically, in the intermediate case 12 where $\sigma_{ov}=1.1$,  $\tilde{E}_{MC}(r)$  decreases in the RZ but the damping is slower compared with   cases 3 and  5 where the amplitude of $\tilde{E}_{MC}(r)$ becomes weaker more rapidly below the base of the CZ. In contrast, both cases 9 and 10, which have $\sigma_{ov}<1$, possess a larger kinetic energy in the meridional flows in the RZ as well as a profile that gradually decays below $r_c$ but is not changing by much overall.  This is similar to what we qualitatively observed in Figure \ref{fig:MCflows}.

We can use these profiles to develop a more quantitative description of the penetration lengthscale of the meridional flows in the RZ.  For each simulation, we fit an exponential function  of the form 
\begin{equation}
\label{eq:fitKE}
f(r)=A\cdot\exp(-\lambda(r-r_c)) .
\end{equation}  to the $\tilde{E}_{MC}(r)$ data within the region spanning from the base of the CZ ($r=r_c$) down to the point where the profile of $\tilde{E}_{MC}$ is well fit by Equation (\ref{eq:fitKE})  as depicted in Figure \ref{fig:KE_MC}b  for case 10.  Once we have measured the decay rate $\lambda$ for each case,  we define the characteristic penetration lengthscale of the meridional flow as $\delta_{MC}=\delta_{MCf}/r_o=2/\lambda$. We note that the prefactor 2 comes from the fact that we fit the exponential function to the kinetic energy of the meridional flows, but we are interested in the lengthscale associated with the meridional flows themselves, namely  $\sqrt{\tilde{E}_{MC}}\propto e^{(-\lambda(r-r_c))^{1/2}}=e^{(-\lambda(r-r_c)/2)}$.

\begin{figure}[ht!]
     \centering
     \includegraphics[scale=.4]{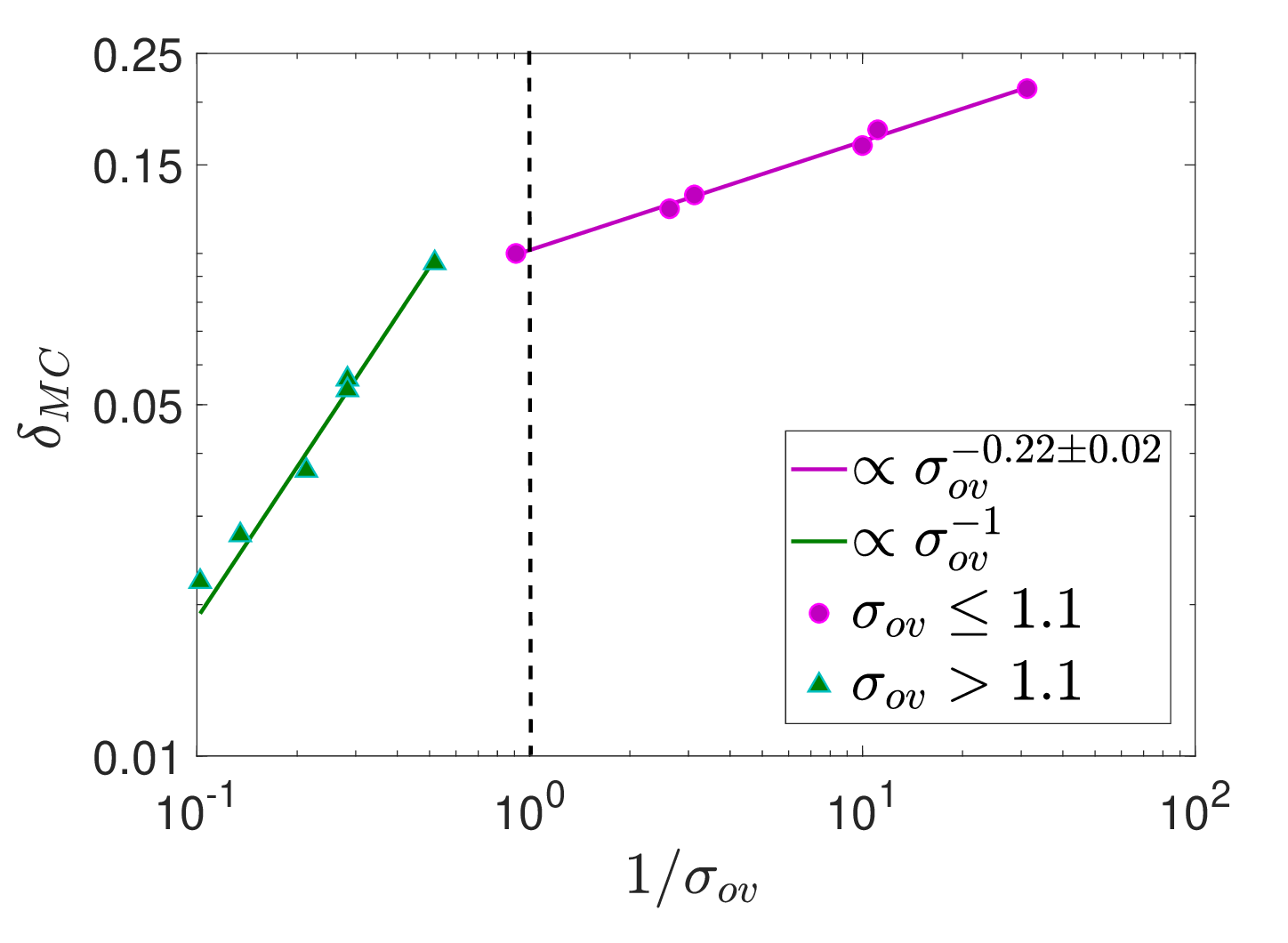}
     \caption{Dependence of $\delta_{MC}$ on the parameter $\sigma_{ov}$. The penetration lengthscale of the meridional flows below the base of the CZ is $\delta_{MC} = (0.1\pm 0.006)\sigma_{ov}^{-0.22\pm 0.02}$ for the values of $\sigma_{ov}\lesssim 1.1$. This scaling does not hold for the runs with $\sigma_{ov}>>1$ whereby $\delta_{MC}\propto \sigma_{ov}^{-1}=(0.19\pm 0.01)\sigma_{ov}^{-1}$.}
     \label{fig:sigma_scale}
 \end{figure}

The relationship between the meridional penetration lengthscale $\delta_{MC}$ and the parameter $\sigma_{ov}$ is illustrated in Figure \ref{fig:sigma_scale}.  We find that this relationship is well described by a broken power law.  For those models with $\sigma_{ov}\lesssim 1.1$, we find that $\delta_{MC}\propto\sigma_{ov}^{-0.22}$.  This scaling law suggests a weak scaling of the penetration lengthscale of the meridional circulation with respect to $\sigma_{ov}$ for values of $\sigma_{ov}\lesssim 1.1$. It also provides a predictive tool that could be used to estimate the penetration depth of the meridional circulation in stellar radiative zones that have  $\sigma<1$ (see Section \ref{sec:disc}).  For values of $\sigma_{ov}>>$ 1.1, we find that $\delta_{MC}\propto\sigma_{ov}^{-1}$.  This latter result is similar to what has been shown in e.g. \cite{GA2009} and \cite{WB12}, associated with the viscous damping of the meridional circulation below the base of the CZ when $\sigma>>1$.


\subsection{Angular Momentum Transport}
\label{sec:AMT}
As discussed in Section \ref{sec:intro}, meridional flows can be driven in response to the redistribution of angular momentum, an effect known as gyroscopic pumping.  Here we explore this effect by examining the balances struck in our models between angular momentum transport due to meridional advection, convective transport, and viscous stresses.  We can obtain an expression for the conservation of angular momentum by taking the zonal component $\phi$ of the momentum equation, multiplying it by the momentum arm $\mu=r\sin\theta$, averaging over both time and longitude and finally assuming a steady state.  When contributions due to the Lorentz force are neglected, we arrive at
\begin{equation}
\label{eq:CAM}
-\bar{\rho}\langle \vel_{mc}\rangle\cdot\nabla\mathcal{L}-(\nabla\cdot \boldsymbol{F}_{RS}+\nabla\cdot \boldsymbol{F}_{VS})=0,
\end{equation}
where $\vel_{mc}=(u_r, u_{\theta})$ and where $\mathcal{L}=\mu^2\Omega$, which can be nondimensionally expressed as
\begin{equation}
\mathcal{L}=r\sin\theta\left(\dfrac{r\sin\theta}{\rm Ek}+\langle u_{\phi}\rangle\right).
\end{equation} 
The  nondimensional transport of angular momentum due to Reynolds stresses and due to viscous stresses is given respectively by
{\color{black}
\begin{equation}
\label{eq:Frs}
\boldsymbol{F}_{RS}=\bar{\rho}r\sin\theta(\langle u_{f,r} u_{f,\phi}\rangle,\langle u_{f,\theta} u_{f,\phi}\rangle),
\end{equation}
\begin{equation}
\label{eq:Fvs}
\boldsymbol{F}_{VS}=\bar{\rho}\left(\langle {u}_{m,\phi}\rangle\sin\theta-r\sin\theta\dfrac{\partial \langle {u}_{m,\phi}\rangle}{\partial r}, \\
\langle {u}_{m,\phi}\rangle\cos\theta-\sin\theta\dfrac{\partial\langle {u}_{m,\phi}\rangle}{\partial\theta}\right).
\end{equation}
}
We expect that in the thermally-relaxed, statistically-steady state, the time-averaged term associated with the Coriolis force   $\bar{\rho}\langle \vel_{mc}\rangle\cdot\nabla\mathcal{L}$ will be balanced primarily by the Reynolds stresses $-\nabla\cdot \boldsymbol{F}_{RS}$ in the convection zone when viscous stresses can largely be neglected, as is typically true for high values of Ra \citep[see e.g.,][]{Featherstone16a}.  Deeper in the stable region, where  the Reynolds stresses are weaker, we expect the viscous stresses  $-\nabla\cdot \boldsymbol{F}_{VS}$ to contribute significantly to the angular momentum balance.

\begin{figure*}[ht!]
     \centering
     \includegraphics[scale=.75]{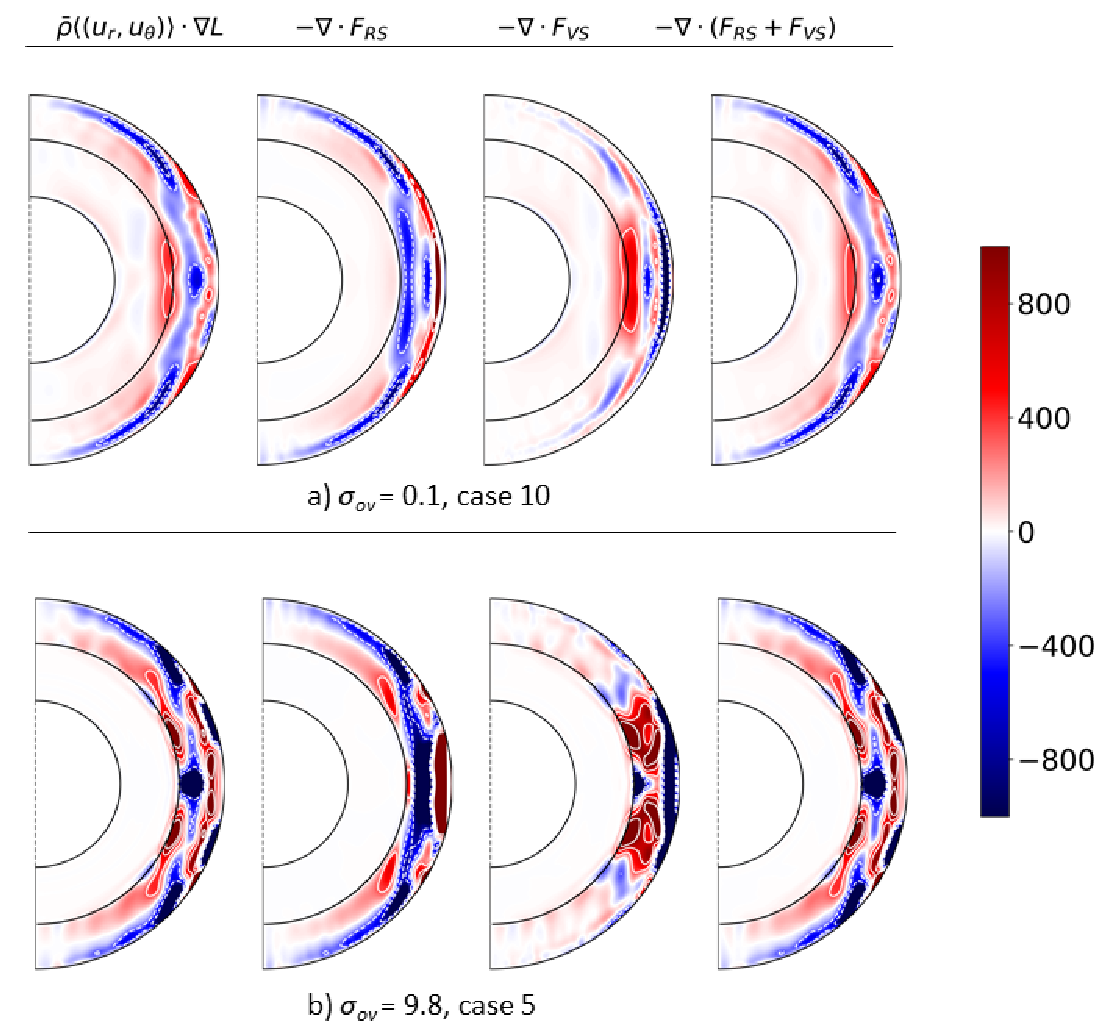}
     \caption{Time- and azimuthally-averaged gyroscopic pumping terms of Equation (\ref{eq:CAM}) for two  $\sigma_{ov}$ cases at the same Ra. The gyroscopic pumping balance is achieved in both cases.   a) In case 10 with $\sigma_{ov}=0.1$, the meridional flows penetrate deeply into the RZ. The gyroscopic pumping occurs across the whole shell but is more prominent within the CZ. b) In case 5 with $\sigma_{ov}=9.8$, the mean flows are confined to a small region below the base of the CZ. The gyroscopic pumping balance holds everywhere across the shell for this run as well, however it is negligible below the base of the CZ where the meridional flows rapidly decay.}
     \label{fig:GP}
 \end{figure*}

For all models  presented in Table \ref{tab:table}, we have verified that in the thermally-relaxed and statistically stationary state, $\bar{\rho}\langle \vel_{mc}\rangle\cdot\nabla\mathcal{L}\approx -(\nabla\cdot \boldsymbol{F}_{RS}+\nabla\cdot \boldsymbol{F}_{VS})$ within the CZ and the RZ. 
In all cases, we notice that $-\nabla\cdot \boldsymbol{F}_{RS}$ is more  dominant  in the bulk of the CZ while $-\nabla\cdot \boldsymbol{F}_{VS}$ is larger within the equatorial region, especially near and somewhat below  the bottom of the CZ.
These results are illustrated in Figure \ref{fig:GP}, where we show the balances achieved in two extreme cases, with $\sigma_{ov}=0.1$ and  $\sigma_{ov}=9.8$. We observe that in the CZ, the balance is achieved between the fluxes associated with both the Reynolds and viscous stresses and the Coriolis force. The Reynolds stresses are somewhat stronger than the  viscous stresses  as they seem to span the whole CZ unlike the viscous stresses  which are mostly present near the equatorial region. This is a result of the level of turbulence dictated by the chosen Ra of our runs shown here, namely,  even higher Ra simulations (which are harder to achieve  computationally in our two-zone spherical shell) need to be considered for the resulting  Reynolds stresses to completely balance the Coriolis term and the viscous stresses to be negligible  across the whole CZ. 
Nonetheless, deeper in the RZ (below the overshoot region), where convective motions are much weaker or non-existent, viscosity becomes more important and counteracts  $\bar{\rho}\langle \vel_{mc}\rangle\cdot\nabla\mathcal{L}$. That is in fact expected in these simulations where there is no magnetism and reinforces  what has already been suggested in the literature, namely that a magnetic field, either primordial or dynamo in nature, needs to be accounted for in solar-like simulations for the dynamics to be more consistent with the Sun \citep[see, e.g.][]{GM98, AG2013,WB18, Matilsky22}.

\begin{figure*}[ht!]
     \centering
     \includegraphics[scale=.55]{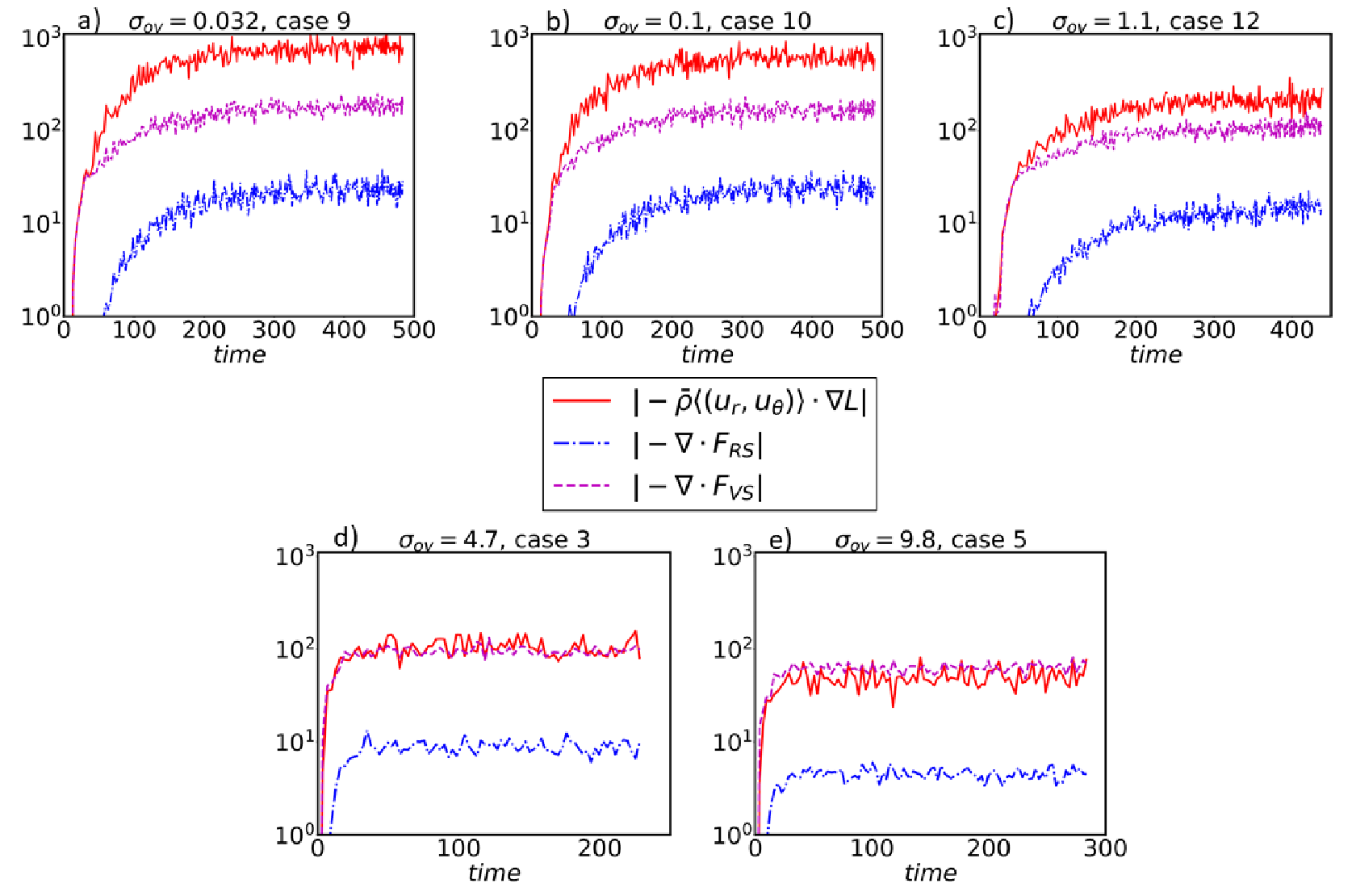}
     \caption{Time evolution of the absolute value of the terms in Equation (\ref{eq:CAM}) volume-averaged from  their overshoot region down to the inner RZ boundary for increasing values of $\sigma_{ov}$ spanning values from a) $\sigma_{ov}=0.032$ to e) $\sigma_{ov}=9.8$. For smaller values of $\sigma_{ov}$, the angular momentum transport is dictated by the flux associated with the Coriolis force, however as $\sigma_{ov}$ increases, the Coriolis term becomes weaker and viscosity plays a more important role and eventually dictates the angular momentum transport in the long term.}
     \label{fig:AMfluxes}
 \end{figure*}

We are interested in understanding the role that each one of the terms in Equation (\ref{eq:CAM}) plays in  the angular momentum transport over time, hence we now investigate  the time evolution of their absolute value. We expect to notice differences in the amplitudes of the fluxes that will depend on $\sigma_{ov}$ for  the cases demonstrated in Figure \ref{fig:AMfluxes} which are all run at the same Ra. That is due to the fact that in the solar-like regime where $\sigma_{ov}\lesssim 1.1$, the meridional flows have a strong presence deeper in the RZ below the overshoot region while for cases with $\sigma_{ov}>>1$, the mean flows are mostly confined to a small region below the base of the CZ and their amplitude decays much faster beyond the convective boundary. 

To examine  the angular momentum transport more quantitatively and understand the  dependence of the fluxes on $\sigma_{ov}$, in Figure \ref{fig:AMfluxes}, we present the profiles of the absolute values of the fluxes, namely   $|-\bar{\rho}\langle \vel_{mc}\rangle\cdot\nabla\mathcal{L}|$, $|-\nabla\cdot \boldsymbol{F}_{RS}|$ and $|-\nabla\cdot \boldsymbol{F}_{VS}|$, volume-averaged from  their corresponding overshoot region  down to the bottom of the RZ  (same calculation as in Equation (\ref{eq:volCZ}) but radially integrated from  $r_i$ to $r_c-\delta_G r_o$). 
We find that   
$|-\bar{\rho}\langle \vel_{mc}\rangle\cdot\nabla\mathcal{L}|$ is much larger than $|-\nabla\cdot \boldsymbol{F}_{VS}|$ for cases 9 and 10 with $\sigma_{ov}<<1$ while its amplitude gradually decreases as $\sigma_{ov}$ becomes larger from panel a) to panel c). In Figure \ref{fig:AMfluxes}d, for case 3,  $|-\bar{\rho}\langle \vel_{mc}\rangle\cdot\nabla\mathcal{L}|\approx|-\nabla\cdot \boldsymbol{F}_{VS}|$, while for case 5 where $\sigma_{ov}>>1$ (Figure \ref{fig:AMfluxes}e), the flux associated with the Coriolis term is now smaller than the flux associated with viscosity, namely $|-\bar{\rho}\langle \vel_{mc}\rangle\cdot\nabla\mathcal{L}|<|-\nabla\cdot \boldsymbol{F}_{VS}|$ which indicates that the angular momentum transport is ultimately dictated by viscosity  unlike what is expected to occur in the solar interior. Finally, we notice that as expected (and discussed above), the Reynolds stresses are  very weak deeper in the RZ below the overshoot region.

These findings are consistent with the differential rotation profiles observed in Figure \ref{fig:diffrot}. For the cases with lower $\sigma_{ov}$ values, we noticed that the differential rotation is more confined within the equatorial region compared with the high-$\sigma_{ov}$ cases. In Figure \ref{fig:AMfluxes}, we found that in the low-$\sigma_{ov}$ cases, the Coriolis term is dominant in the angular momentum transport in the RZ while as $\sigma_{ov}$ becomes larger, the flux related to the mean flows decreases and as a result the viscous stresses  eventually control angular momentum transport. Consequently, in the cases with low $\sigma_{ov}$, the propagation of the differential rotation  in the RZ is due to the advection of the mean flows while in the cases with high $\sigma_{ov}$, it is due to viscous diffusion.

\begin{figure*}[ht!]
     \centering
     \includegraphics[scale=.55]{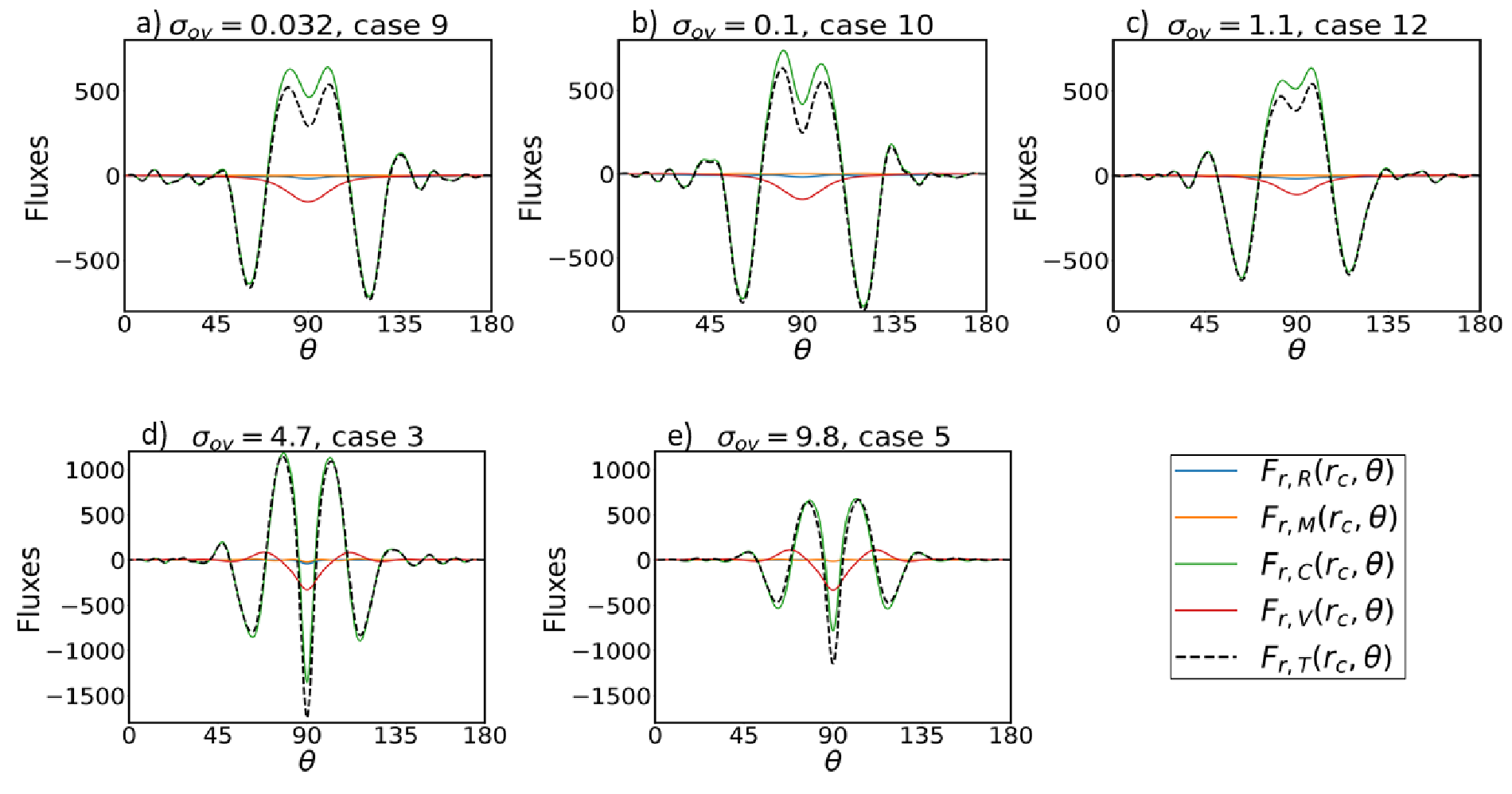}
     \caption{Profiles of the time- and azimuthally-averaged fluxes given in Equations (\ref{eq:termsAMr1})-(\ref{eq:termsAMtot}) computed at the base of the CZ for increasing values of $\sigma_{ov}$ spanning values from a) $\sigma_{ov}=0.032$ to e) $\sigma_{ov}=9.8$. For smaller (larger) values of $\sigma_{ov}$, both $F_{r,T}$ and $F_{r,C}$ are positive (negative) near the equator indicating that the angular momentum transport is outward (inward) there.}
     \label{fig:fluxes}
 \end{figure*}

To better understand this behavior, in Figure \ref{fig:fluxes}, we  show the the time- and azimuthally-averaged radial fluxes related to the Reynolds stresses, the mean flows, the Coriolis force, and the viscous stresses as well as the total radial flux, all plotted  at the base of the CZ (at $r=r_c$) and given respectively by

\begin{eqnarray}
\label{eq:termsAMr1}
F_{r,R}(r_c,\theta)=\langle\bar{\rho}r \sin\theta({u_{f,r} u_{f,\phi}})\rangle|_{r_c},\\
\label{eq:termsAMr2}
F_{r,M}(r_c,\theta)=\langle\bar{\rho}r \sin\theta({u}_{m,r} {u}_{m,\phi})\rangle|_{r_c},\\
\label{eq:termsAMr3}
F_{r,C}(r_c,\theta)=\left\langle\bar{\rho}r \sin\theta\left(\dfrac{{u}_{m,r} r\sin\theta}{{\rm Ek}}\right)\right\rangle|_{r_c},\\
\label{eq:termsAMr4}
F_{r,V}(r_c,\theta)=\left\langle\bar{\rho}r \sin\theta\left(-r\dfrac{\partial}{\partial r}\left(\dfrac{{u}_{m,\phi}}{r}\right) \right)\right\rangle|_{r_c},\\
\label{eq:termsAMtot}
F_{r,T}(r_c,\theta)=F_{r,R}(r_c,\theta)+F_{r,M}(r_c,\theta)+F_{r,C}(r_c,\theta)+F_{r,V}(r_c,\theta).
\end{eqnarray}

 We  notice that in the lower $\sigma_{ov}$ cases, the total flux $F_{r,T}$ at the base of the CZ is positive near the equatorial region, similarly to the flux related to the Coriolis force $F_{r,C}$, while the viscous flux $F_{r,V}<0$ there. On the other hand, in the cases with $\sigma_{ov}>>1$, $F_{r,T}$ is negative around the equator following the behavior of $F_{r,C}$ and $F_{r,V}$. This finding suggests that the suppression of the propagation of the differential rotation within the equator for the low-$\sigma_{ov}$ runs is a result of the radially outward flux of angular momentum, which derives primarily from the interaction of meridional flow with the Coriolis force.  This lies in contrast with the high-$\sigma_{ov}$ cases where $F_{r,T}<<0$ leads to the inward angular momentum transport and the deepening of the differential rotation in the RZ. 
 
 We note that the amplitudes and profiles of these fluxes do not only depend on $\sigma(r)$ but also on Ra and the density stratification $N_{\rho}$. However, we found that in all of our runs, the differential rotation for the lower $\sigma_{ov}$ runs is more contained within the equator and that $F_{r,C}>0$ there, unlike in the higher $\sigma_{ov}$ cases where both $F_{r,C}$ and $F_{r,T}$ are negative.

\section{Conclusions}
\label{sec:disc}
\subsection{Summary and Discussion of Our Results}
{\color{black} Our aim in this work has been to examine the dynamical balances associated with axisymmetric mean flows in the radiative interior of a low-mass star such as the Sun.   As discussed in \citet{WB12}, in the absence of magnetism, the balance that expresses the relative importance of advective and viscous transport in the radiative interior is well-characterized by a single parameter, $\sigma$ (Eq. \ref{eq:sigmaeq}). 
Viscous effects are negligible when $\sigma<1$, as is appropriate for a stellar interior, and are dominant when $\sigma>1$. 

To date, models in the stellar-relevant regime have only been run in  axisymmetric spherical or Cartesian geometries, which cannot account for latitudinal variations of the mean flows.  Spherical models incorporating a region of overshoot have in turn been run only in the $\sigma>1$ regime.  In order to bridge this gap, we have analyzed the results from a series of 3D global simulations spanning both the $\sigma\leq 1$ and $\sigma>1$ regimes.  These models are constructed in spherical geometry, incorporate rotation, and consist of  an outer convective region overlying a stably stratified layer.

Much as in the models of \citet{GA2009}, \citet{WB12}, and \citet{AG2013}, we find that in the $\sigma<1$ regime appropriate for the Sun, meridional flows can penetrate deeply into the RZ.  In the $\sigma>1$ regime, they decay exponentially in depth with a lengthscale proportional to $1/\sigma$ (see Fig. \ref{fig:sigma_scale}).  The deeply- and shallowly-penetrating flows arising in these two regimes lead to markedly different balances in the angular momentum transport. For increasing values of $\sigma_{ov}$, the angular momentum flux arising from the Coriolis force becomes weaker, and for cases with $\sigma_{ov}>>1$, the angular momentum transport is dictated primarily by the viscous flux in the long term.  In contrast, for simulations with $\sigma_{ov}<<1$, the flux associated with the Coriolis term is much larger than the viscous flux indicating that in the solar-like regime, the mean flows are responsible for the transport of angular momentum.

The dominance of the Coriolis term in the $\sigma_{ov} <<1$ regime is particularly notable near the equator and leads to differences in the differential rotation that develops below the convection zone.  The differential rotation established in all models consists of a rapidly-rotating equator and slowly-rotating polar regions.  This general profile was found to extend below the base of the CZ for all models due to either viscous diffusion ($\sigma_{ov}>>1$) or advection by the mean flows ($\sigma_{ov}\leq 1.1$).  We note, however, that for models with $\sigma_{ov}<1$, the rapid equatorial differential rotation does not extend as deeply as it does in those with  $\sigma_{ov}>1$ (see Fig. \ref{fig:diffrot}).  These results suggest that meridional flows may play a role in tachocline confinement at low latitudes, though a definitive statement cannot be made here owing to the fact that this study did not consider the effects of magnetism.}

Currently, even the most state-of-the-art simulations cannot capture real solar (and stellar) values due to computational constraints arising from the required spatio/temporal resolution. As such, it is important to ensure that since solar values are unattainable, they should not be used for certain input parameters (e.g. $N$, and $\Omega$) and ignored for others (e.g. Pr) as this may lead to non-solar/stellar parameter regimes. Consequently, the resulting dynamics cannot effectively capture the dynamical processes taking place in the solar interior and at the same time, any extrapolations to solar/stellar parameters through derived scaling laws are dubious when they result from simulations that operate in non-solar/stellar regimes.

\subsection{Implications for Dynamical Processes in the Solar Interior}
As we showed in Section \ref{sec:model}, the solar $\sigma$ profile is smaller than one in the upper radiative zone and approximately equal to 0.5 around the base of the solar tachocline. Since we are reporting our results based on the location of the overshoot lengthscale, we will use our estimate of the solar overshoot depth from \cite{KF21} and attempt to estimate the penetration lengthscale of the meridional flows in the solar radiative interior. We have to note, however, that even though our scaling law is derived from our runs that were operated within the solar-like regime where $\sigma<1$, we still need to be cautious when extrapolating our numerical results to the Sun and stars in general. That is due to the fact that most of the other parameters have non-solar values, for instance our largest Ra $=10^6$ while Ra$_{\odot}=10^{20}$ and we used Pr $=1$ while Pr$_{\odot}=10^{-6}$ \citep[e.g.,][]{Ossendrijver2003}.

With that in mind, we may proceed to provide an estimate of $\delta_{MC,{\odot}}$. 
In \cite{KF21}, we assessed that the solar overshoot lengthscale is approximately equal to $0.1H_p$, where $H_p=5.7e9$ cm is the pressure scale-height measured at the base of the solar CZ (using Model S). Then the location of the overshoot depth is at $r_{ov}\approx 0.7r_{\odot}$ hence $\sigma_{ov,{\odot}}\approx 0.42$ (where $r_{\odot}=0.9983R_{\odot}$ is the radius at the base of the granulation layer and $R_{\odot}=6.957e10$ cm). From the scaling law derived in Section \ref{sec:results}, we can then obtain $\delta_{MC,\odot}=0.1r_{\odot}/(\sigma_{ov,\odot}^{0.22})\approx 0.12r_{\odot}$ which corresponds to $\delta_{MC,\odot}\approx 1.46 H_p$. This result indicates that the meridional circulation might be penetrating much more deeply in the solar interior than accounted for or anticipated by previous numerical studies and observations. In fact, we find that the meridional flows might travel as far as down to $r=0.59r_{\odot}=4.098e10$ cm, which is even below the overshoot region and the tachocline. 

If that is indeed the case, the meridional circulation may have a greater role to play in the transport of chemical species within the RZ, as well as in the dynamical balances taking place within and beyond the tachocline region. For instance, \citet{GM98} suggested that a primordial magnetic field in the RZ could confine the solar tachocline and enforce uniform rotation there by interacting with the large-scale mean flows gyroscopically-pumped from the CZ deeper into the RZ in a way that they could halt the magnetic field from entering the CZ. Also, our results may hold implications for flux transport models where, in tandem with the tachocline, the meridional flow plays an important role in modulating the magnetic cycle \citep[e.g., ][]{Dikpati1999,Charbonneau2010}.  We plan to explore these topics in a future study that incorporates magnetism into the models presented here.
\\\\

 \noindent We thank Brad Hindman and Bhishek Manek for helpful discussions. L.K. acknowledges support from NASA's Early Career Investigator Program grant No. 80NSSC21K0455.  N.F. and L.K were both supported by NASA grants  80NSSC20K0193 and 80NSSC24K0125. N.F. acknowledges additional funding from NASA grant No.  80NSSC22M0162. The authors acknowledge the NASA
High-End-Computing (HEC) program for providing the computational
resources essential for conducting the numerical simulations of this work. This work utilized resources from the University of Colorado Boulder Research Computing Group, which is supported by the National Science Foundation (awards ACI-1532235 and ACI-1532236), the University of
Colorado Boulder, and Colorado State University. The
Rayleigh code has been developed with support by the
National Science Foundation through the Computational
Infrastructure for Geodynamics under grants NSF-0949446
and NSF-1550901.

\bibliography{biblio}{}
\bibliographystyle{aasjournal}

\end{document}